\documentclass[10pt,conference]{IEEEtran}

\usepackage{cite}
\usepackage{amsmath,amssymb,amsfonts}
\usepackage{algorithmic}
\usepackage{graphicx}
\usepackage{textcomp}
\usepackage[colorlinks=false,hidelinks]{hyperref}
\usepackage[capitalise,nameinlink,noabbrev]{cleveref}
\usepackage{xcolor}
\usepackage{threeparttable}
\usepackage{multirow}
\usepackage{array}
\usepackage{makecell}
\usepackage{calc}

\usepackage[square,numbers,sort&compress]{natbib}

\usepackage[framemethod=TikZ]{mdframed}
\usepackage{xpatch}

\makeatletter
\xpatchcmd{\endmdframed}
  {\aftergroup\endmdf@trivlist\color@endgroup}
  {\endmdf@trivlist\color@endgroup\@doendpe}
  {}{}
\makeatother
    
\begin{document}

\bstctlcite{IEEEexample:BSTcontrol}

\title{Harnessing Patterns to Support the Development \\of Hybrid Quantum Applications}

\author{
\IEEEauthorblockN{Daniel Vietz, Martin Beisel, Johanna Barzen, Frank Leymann, Lavinia Stiliadou, Benjamin Weder} 
\IEEEauthorblockA{\textit{University of Stuttgart,}
\textit{Institute of Architecture of Application Systems, Universitätsstrasse 38, 70569 Stuttgart, Germany} \\
\textit{\{firstname.lastname\}@iaas.uni-stuttgart.de}}
}

\maketitle

\begin{abstract}
Quantum computing provides computational advantages in various domains.
To benefit from these advantages complex hybrid quantum applications must be built, which comprise both quantum and classical programs.
Engineering these applications requires immense expertise in physics, mathematics, and software engineering.
To facilitate the development of quantum applications, a corresponding quantum computing pattern language providing proven solutions to recurring problems has been presented.
However, identifying suitable patterns for tackling a specific application scenario and subsequently combining them in an application is a time-consuming manual task.  
To overcome this issue, we present an approach that enables (i)~the automated detection of patterns solving a given problem, (ii)~the selection of suitable implementations fulfilling non-functional requirements of the user, and (iii)~the automated aggregation of these solutions into an executable quantum application.
\end{abstract}


\section{Introduction}
\label{sec:introduction}
\noindent
Given the rapid developments in the quantum computing domain, building qualitative quantum applications is of vital importance to efficiently solve problems in various application areas, e.g., chemistry, finance, or scientific
simulations~\cite{national2019quantum, cao2018qcchemistry}.
However, quantum applications are typically hybrid, which complicates their development process, as both quantum and classical programs must be implemented and integrated~\cite{Weder2021_OrchestrationsInSuperposition,Beisel2023_PracticalQCIntroduction}.
Therefore, quantum application developers must possess expertise in different fields, such as physics, mathematics, and software engineering~\cite{nielsen2002quantum, weder2022SEDevLifecycle}.

A well-known concept to capture proven solutions for recurring problems in an abstract manner is the documentation technique of \textit{patterns}~\cite{Alexander1977_PatternLanguage}.
In software engineering, patterns are commonly used as a basis for design decisions when developing applications~\cite{Gamma1995_PatternLanguages}.
Patterns have a well-defined structure, and each pattern is related to other patterns with a defined semantic.
For example, relations can indicate that two patterns are alternatives to each other or that applying the first pattern also requires using another.
The interconnected set of all patterns of a certain domain form a so-called \textit{pattern language}~\cite{Alexander1977_PatternLanguage}.
For example, the quantum computing pattern language~\cite{Leymann2019_QuantumPatternLanguage} captures important patterns and their relations to support developers in building quantum applications.\looseness=-1

However, when building quantum applications, identifying the set of relevant patterns is a complex task that requires an understanding of the patterns within the pattern language as well as their relations.
As a starting point for identifying all patterns relevant to solving the problem at hand, so-called entry point patterns must be identified~\cite{Reinfurt2019_EntryPoints}.
A path comprising all relevant patterns can be found using the semantic connections of these entry point patterns to related patterns.
Since patterns contain abstract descriptions of proven solutions, building an application for a use case is still complex and time-consuming.
To tackle this problem, the concept of concrete solutions, i.e., implementations of a pattern for a specific scenario, has been introduced~\cite{Falkenthal2017_SolutionLanguages}.
However, to successfully build a quantum application by reusing concrete solutions, they need to be analyzed for their suitability and subsequently aggregated.\looseness=-1

To overcome these issues, in this paper, we introduce an approach that (i)~enables the automated detection of entry point patterns as well as the identification of related patterns required for solving use-case-specific problems in the quantum domain based on the textual input provided by a user.
Further, we (ii)~select suitable concrete solutions implementing the required functionalities based on the previously identified patterns as well as the non-functional requirements extracted from the user input.  
Finally, we (iii)~automatically aggregate the selected concrete solutions into a deployable and executable quantum application.
To validate the practical feasibility of our approach, we present a system architecture, a corresponding prototypical implementation, and a case study realizing a typical quantum application using our approach.\looseness=-1

The remainder of the paper is as follows:
\Cref{sec:fundamentals} presents fundamentals and discusses our problem statement.
In \Cref{sec:approach}, we introduce our approach for the automated generation of hybrid quantum applications based on the quantum computing patterns.
\Cref{sec:arch} showcases the system architecture realizing our approach and its prototypical implementation.
In \Cref{sec:casestudy}, we present a case study generating a quantum application for a typical use case.
Finally, \Cref{sec:relatedWork} discusses related work, and \Cref{sec:fw} concludes the paper.\looseness=-1

\section{Fundamentals \& Problem Statement}
\label{sec:fundamentals}
\noindent
In this section, fundamentals about pattern and solution languages, as well as the quantum computing patterns are introduced.
Furthermore, we present our research question.\looseness=-1

\begin{figure*}[!ht]
\includegraphics[width=1\textwidth,trim=0 0 0 0 ,clip]{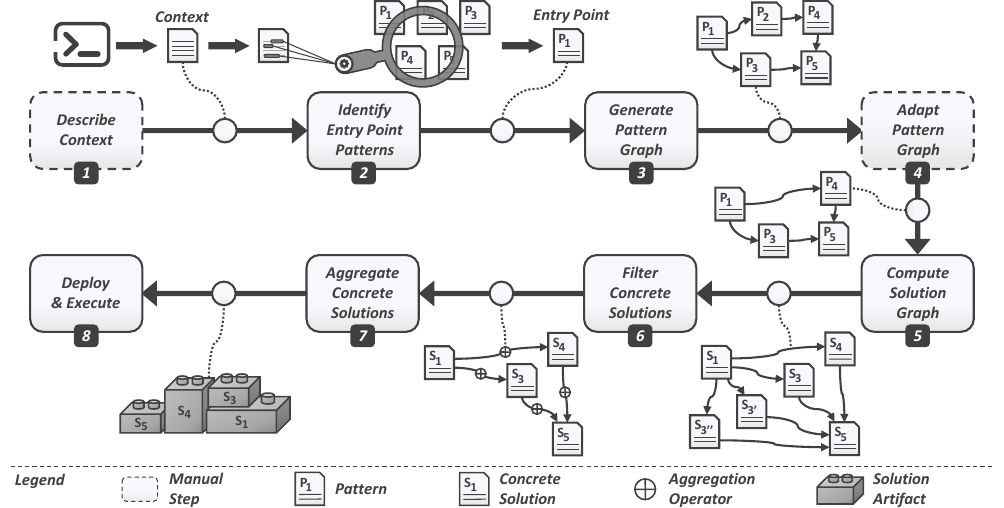}
\caption{Overview of the pattern-based approach for building and executing hybrid quantum applications}
\label{fig:approach}
\end{figure*}

\subsection{Patterns \& Solution Languages}
\noindent
Patterns are used in many domains, such as computer science and architecture, to describe proven solutions to recurring problems in a structured manner~\cite{Alexander1977_PatternLanguage,Hohpe2003_PatternLanguages}.
The description of each pattern follows a well-defined format, e.g., specifying the context in which the pattern can be applied, the problem it solves, or the corresponding solution to apply~\cite{Gamma1995_PatternLanguages}.
Furthermore, each pattern is semantically connected to all related patterns, forming a so-called pattern language~\cite{Alexander1977_PatternLanguage}.

To apply abstract solutions described by patterns, corresponding software artifacts are required, implementing their functionality for a specific scenario~\cite{Falkenthal2019_SolutionAggregation}.
These software artifacts are referred to as \textit{concrete solutions} and are organized in a solution language~\cite{Falkenthal2017_SolutionLanguages}.
Concrete solutions are linked to the patterns of the corresponding pattern language.
To solve a larger problem requiring multiple patterns, concrete solutions for a set of patterns can be aggregated~\cite{Falkenthal2019_SolutionAggregation}.

\subsection{Quantum Computing Patterns}
\noindent
A pattern language specifically focussing on the quantum computing domain was introduced by Leymann~\cite{Leymann2019_QuantumPatternLanguage} and is continuously extended~\cite{Beisel2022_ErrorHandlingPatterns,Bechtold2023_CuttingPatterns,Georg2023_PatternsQuantumExecution,Stiliadou2025_QMLPatterns,Beiselini2025_OpsPattern}.
The \textit{quantum computing pattern language} aims to support quantum software engineers in developing, operating, and adapting hybrid quantum applications.
Thereby, it covers different aspects and phases of the lifecycle for hybrid quantum applications~\cite{weder2022SEDevLifecycle}:
For example, the \textit{error handling patterns}~\cite{Beisel2022_ErrorHandlingPatterns} summarize concepts to mitigate and correct computational errors occurring due to the noisyness of today's quantum computers.
The \textit{circuit cutting patterns}~\cite{Bechtold2023_CuttingPatterns} describe another approach to cope with noisy quantum computers by showcasing how large quantum circuits can be cut into smaller quantum circuits which can be executed with higher accuracy.
Furthermore, the \textit{execution patterns}~\cite{Georg2023_PatternsQuantumExecution} discuss different strategies for deploying and executing quantum circuits as well as quantum applications.\looseness=-1

\subsection{Problem Statement \& Research Question}
\noindent
While the quantum computing patterns support quantum software engineers during the development and execution process of hybrid quantum applications, it remains difficult to identify the required set of patterns needed for a certain application.
Thus, either these patterns have to be identified by an expert knowing all patterns of the pattern language, or this results in a time-consuming task for the developers~\cite{Reinfurt2019_EntryPoints}.
Furthermore, the patterns provide abstract solutions to various problems and require the identification or development of a corresponding implementation.
Finally, the implementations for the patterns must be combined which can be a complex and error-prone task, especially if the implementations are heterogeneous, e.g., using different programming languages and data formats~\cite{weder2022SEDevLifecycle,Falkenthal2019_SolutionAggregation}.
This leads us to our Research Question~(RQ):

\begin{mdframed}[skipabove=4pt,skipbelow=0.5pt,innerbottommargin=2.5px,innertopmargin=3.5px\baselineskip, innerleftmargin=8px, innerrightmargin=8px]
\hspace{-5px}
\textit{\textbf{RQ:} ``How can suitable patterns be identified based on textual input and how can they be used for a semi-automatic generation of hybrid quantum applications?''}
\end{mdframed}

\section{Building Hybrid Quantum Applications \\Using Patterns}
\label{sec:approach}
\noindent
In this section, we discuss our approach for the pattern-based development and execution of quantum applications.
\Cref{fig:approach} depicts its eight phases, which are discussed in the following.\looseness=-1

\subsection{Describe Context}
\noindent
First, the user specifies the context in which the hybrid quantum application should be developed.
This comprises describing the problem that has to be solved by the application or the needed functionality, i.e., the functional requirements.
Further, the non-functional requirements can be defined, e.g., expected runtime or the preferences for the deployment of the application.
To reduce the complexity for the user, the context is specified utilizing a textual description, which is automatically analyzed in the next phase of the approach.\looseness=-1

\subsection{Identify Entry Point Patterns}
\noindent
In the second phase, \textit{entry point patterns}~\cite{Reinfurt2019_EntryPoints} are identified, i.e., patterns that can be used as starting points for finding a path through the pattern language that comprises all patterns required to solve a problem.
To find entry point patterns for the textual problem description provided by the user in the first step of our approach, a pre-processing of the description is required.
Users can describe multiple problems in their textual input, e.g., they want to first cluster their input data and subsequently classify it.
Thus, the input has to be split, and each part related to one problem is processed separately.
Since users can provide both functional and non-functional requirements in their descriptions and only the functional requirements are relevant for the identification of suitable patterns, they need to be separated.
The natural language processing capabilities of large language models, such as GPT4~\cite{Achiam2023_Gpt}, can help to extract and separate the functional and non-functional requirements as keywords in the provided context description.
The extracted keywords containing the information about the functional requirements can then be utilized to identify suitable entry point patterns.
For this, the context, problem description, solution, and known uses of each pattern within the pattern language are compared to the extracted keywords, e.g., using similarity measures such as \textit{Cosine Similarity}~\cite{Singhal2001_CosineSimilarity}.
Finally, the different potential entry point patterns are ranked based on the similarity to the extracted keywords and filtered by a threshold defining the minimum required similarity.
Furthermore, for each potential entry point, it is checked if they meet the non-functional requirements, e.g., if the runtime complexity is suitable for the user description.
If no suitable entry point pattern is found, the user is asked to provide additional details about their problem.
Otherwise, the entry point pattern with the highest similarity is used in the next step.\looseness=-1

\subsection{Generate Pattern Graph}
\noindent
Once suitable entry point patterns have been identified, a path through the pattern language has to be found that comprises all patterns relevant to solving the problem at hand.
To identify this path, the interconnections within the pattern language provided through the related patterns section are used.
By navigating through the links within the pattern language, potentially relevant patterns can be found and evaluated regarding their necessity for the given context.
The automated detection of all required patterns is difficult for some patterns, e.g., as their known uses are very generic and difficult to associate with the context description at hand.
Experts can pre-define pattern graphs and directly attach them to the corresponding entry point pattern in an offline pre-processing step.
Thus, if such graphs are available, they can be loaded in this step to improve the quality of the pattern graph generation.\looseness=-1

\subsection{Adapt Pattern Graph}
\noindent
After generating the pattern graph, it can be analyzed and adapted by the user in an optional step.
Thereby, new patterns are added, or contained patterns are removed.
The reason for this can be that a certain pattern has a consequence that is not intended by the user and is thus removed.
For example, a circuit cutting pattern~\cite{Bechtold2023_CuttingPatterns} leads to additional classical overhead during execution that is not required if the pattern is removed.
However, this also affects the functionality of the resulting quantum application, as large quantum circuits might not be successfully executable without circuit cutting.\looseness=-1

\subsection{Compute Solution Graph}
\noindent
In the fifth phase, a solution graph is computed based on the pattern graph~\cite{Falkenthal2019_SolutionAggregation}.
For this, a solution repository is accessed and all available concrete solutions are retrieved for all patterns.
The resulting solution graph comprises all concrete solutions for all utilized patterns, and hence, can comprise multiple concrete solutions for each pattern as depicted in \Cref{fig:approach}.
Thereby, the edges between the concrete solutions are adopted from the corresponding patterns.

\begin{figure*}[!ht]
\includegraphics[width=1\textwidth,trim=0 0 0 0 ,clip]{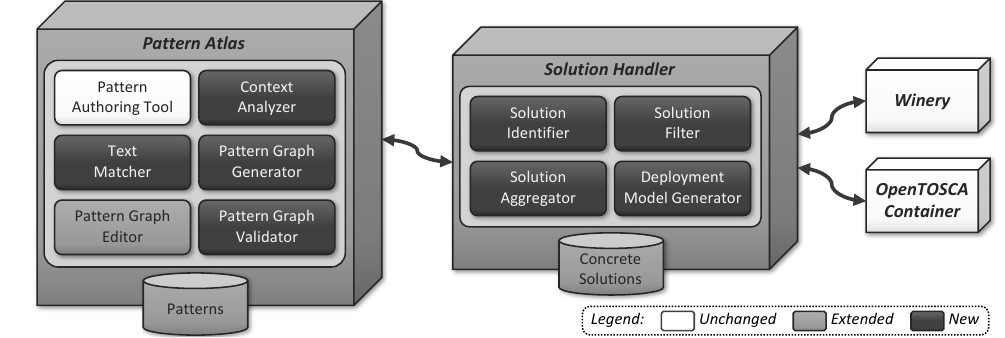}
\caption{System architecture supporting the pattern-based approach for building and executing hybrid quantum applications}
\label{fig:architecture}
\end{figure*}

\subsection{Filter Concrete Solutions}
\noindent
To build the hybrid quantum application, only a single concrete solution per pattern is required.
First, all concrete solutions are filtered based on the non-functional requirements extracted from the user description in the second phase of the approach.
To enable this filtering, the solutions in the repository are annotated with policies describing their non-functional properties.
For example, in \Cref{fig:approach}, three solutions corresponding to pattern P\textsubscript{3} exist.
However, solution S\textsubscript{3'} does not satisfy a non-functional requirement and is thus removed from the graph.
Suitable concrete solutions must also be combinable with the concrete solutions of all connected patterns to finally aggregate them into the overall application.
Thus, so-called \textit{aggregation operators} are searched in the solution repository for each connection and are added to the solution graph.
An aggregation operator specifies how two concrete solutions are combined, e.g., a script defining the point in a program at which a code snippet must be inserted to add certain functionality to the program~\cite{Falkenthal2017_SolutionLanguages}.
If no suitable aggregation operator exists for a connection, the connection is removed from the graph.
A path in the solution graph is considered valid if it comprises concrete solutions for every required pattern in the pattern graph.
If no valid solution path can be found, the method is aborted, and the user is requested for additional input.
If there are multiple valid solution paths, the first one is selected.
In future work, we plan to evaluate how to select the most suitable solution path, e.g., based on collected data about how successfully the contained aggregation operators were applied within past scenarios.\looseness=-1

\subsection{Aggregate Concrete Solutions}
\noindent
In the seventh step, the concrete solutions contained in the computed solution graph are aggregated into a hybrid quantum application.
For this, the aggregation operators defined within the solution graph are applied sequentially starting from the entry point pattern.
As discussed in the second step, the textual description might require multiple algorithms, which also results in multiple entry point patterns and independent solution graphs.
Thus, the solution graphs resulting from the different entry point patterns are aggregated independently and are subsequently combined.
For example, if the solution to the textual description requires applying a clustering algorithm and a classification algorithm, the aggregated solution for the clustering step and the classification step are combined.
All details about aggregating multiple concrete solutions using aggregation operators are discussed by Falkenthal et al.~\cite{Falkenthal2017_SolutionLanguages,Falkenthal2019_SolutionAggregation}.\looseness=-1

\subsection{Deploy \& Execute}
\noindent
In the last phase of the approach, the hybrid quantum application is deployed and executed.
As the automation of steps one to seven frees developers from understanding all peculiarities of the application, e.g., needed dependencies, the manual deployment of the application is complex, error-prone, and time-consuming~\cite{Breitenbuecher2017_DeclarativeVsImperative}.
Thus, the deployment is automated using a deployment system~\cite{weerasiri2017taxonomy}.
For this, the concrete solutions must specify their requirements, which are also stored in the solution repository.
These requirements are used to generate a deployment model, which can be used to automatically deploy the quantum application~\cite{Vietz2022_QuantumScriptSplitting, Weder2023_QuantumWorkflowProvenance,Beiselini2025_OpsPattern}.\looseness=-1

\section{Architecture \& Prototype}
\label{sec:arch}
\noindent
In this section, we present the system architecture realizing our approach and the corresponding prototypical implementation.

\subsection{System Architecture}
\noindent
\Cref{fig:architecture} gives an overview of the system architecture supporting our approach.
It consists of four main components:
(i)~The \textit{Pattern Atlas} enabling the creation and management of patterns,~(ii) the \textit{solution handler} managing and aggregating concrete solutions,~(iii) \textit{Winery} to generate required deployment models, and~(iv) the \textit{OpenTOSCA Container} for executing the generated deployment models.

Pattern Atlas~\cite{Leymann2021_PatternAtlas} is a tool for authoring, managing, and visualizing patterns and pattern languages.
It comprises six components:
The \textit{pattern authoring tool} provides a graphical pattern editor to guide users during the pattern authoring process.
The \textit{context analyzer} is a new component enabling the extraction of the keywords describing the functional and non-functional requirements from the textual input provided by the user.
To identify the entry point patterns, the \textit{text matcher} compares these keywords to the content of the quantum computing patterns. 
Another new component is the \textit{pattern graph generator}, which computes the remaining patterns needed to fulfill the user requirements and arranges them in the pattern graph based on their relations.
This pattern graph can be visualized using the \textit{pattern graph editor}, which was extended to support the manual adaptation of pattern graphs.
The newly added \textit{pattern graph validator} verifies that all changes made by the user result in a valid pattern graph, e.g., by checking if it is a connected graph.
All available \textit{patterns} are stored in a corresponding repository.\looseness=-1

The solution handler is in charge of storing and aggregating concrete solutions to build hybrid quantum applications:
The \textit{solution identifier} provides the functionality to identify the concrete solutions for a given pattern.
Another new component is the \textit{solution filter}, which filters the concrete solutions available for a certain pattern based on different criteria.
For example, the filtering ensures that non-functional requirements, such as security or scalability, are fulfilled by the concrete solutions.
The \textit{solution aggregator} semi-automatically aggregates the concrete solutions from a given solution graph to build the hybrid quantum application.
To automate the deployment of this hybrid quantum application, a corresponding deployment model is generated by the \textit{deployment model generator} based on the requirements of the different aggregated concrete solutions.
The \textit{concrete solutions} are stored in a dedicated repository.\looseness=-1

Winery~\cite{Kopp2013_Winery} is a graphical modeling tool for deployment models based on the TOSCA standard~\cite{TOSCA_Specification}.
It is used by the solution handler to generate the deployment models for the aggregated hybrid quantum applications.
Thereby, it takes the requirements of the user regarding the deployment of the application into account, e.g., that a certain quantum cloud provider should not be used.
Finally, the OpenTOSCA Container~\cite{Binz2013_OpenTOSCA} is a TOSCA-compliant deployment system enabling the automated deployment of the hybrid quantum application based on the previously generated deployment model. \looseness=-1

\subsection{Prototypical Implementation}
\noindent
The prototypical implementation realizing our system architecture is publicly available on GitHub~\cite{ustutt2024useCases}.
All implemented components are provided as open-source projects.
The Pattern Atlas consists of a backend implemented in Java and a frontend realized using the Angular framework.
To perform the AI-based extraction of the functional and non-functional requirements within the textual input provided by the user, ChatGPT is utilized.
The solution handler is a component implemented in Python and utilizes the framework Red Baron to aggregate Python-based solutions into hybrid quantum applications.
To store and identify concrete solutions for a certain pattern, it uses a relational database with corresponding queries.
The deployment model generation is realized by an HTTP connector utilizing the functionality provided by Winery.

\section{Case Study}
\label{sec:casestudy}
\noindent
In the following, we show the practical feasibility of our pattern-based approach by building a hybrid quantum application that solves the 3-SAT problem.
To ensure the reproducibility and reusability of the case study, the required source code and a detailed description of all performed steps are available on GitHub~\cite{ustutt2024useCases}.\looseness=-1

\begin{figure}[t]
    \centering
    \includegraphics[width=\linewidth]{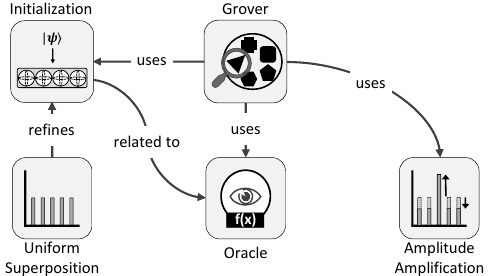}
    \caption{Generated Pattern Graph}
    \label{fig:usecase_patterngraph}
\end{figure}

First, the user describes the problem to solve, its context, and the requirements using the following textual input:
\textit{'Given a set of variables and a boolean logic formula, I need to determine a variable assignment that satisfies the formula, if one exists.
The resulting application should be executed using quantum computers from IBMQ'}.
Based on this input, Grover's algorithm~\cite{grover1996fast} is identified as the entry point pattern, as it enables solving the satisfiability problem.
Then, the complete pattern graph is calculated based on the related patterns of the Grover pattern, as depicted in \Cref{fig:usecase_patterngraph}.
It comprises, e.g., the \textit{initialization} pattern to create the initial \textit{uniform superposition} required by Grover's algorithm or the \textit{oracle} pattern which is used to encode the problem instance to solve.
As the pattern graph comprises all required patterns, it is not manually adapted and the solution graph is computed next.
The solution graph has the same structure as the pattern graph, but for the case study two solutions exist per pattern, one implementing the functionality using Braket that can be executed on AWS quantum computers and one using Qiskit for IBMQ quantum computers.
Due to the user requirement to execute the application using quantum computers from IBMQ, the Braket-based solutions are filtered, resulting in exactly one solution per pattern.
These solutions are automatically combined using the solution handler to build the deployable hybrid quantum application.\looseness=-1

\section{Related Work}
\label{sec:relatedWork}
\noindent
Patterns are used in various approaches to automatically generate or adapt models and applications:
Saatkamp~et~al.~\cite{Saatkamp2019_SolveDetectedProblems} present an approach to detect problems in deployment models and to solve them by applying concrete solutions to adapt the model.
Falkenthal~et~al.~\cite{Falkenthal2017_SolutionLanguages} introduce a concept to transform pattern-based deployment models into executable deployment models using concrete solutions.
Similarly to the concept of concrete solutions, Hallstrom~et~al.~\cite{hallstrom2009reusing} present the concept of design refinements, which enables the specialization of patterns for specific scenarios and requirements. 
Furthermore, their approach enables linking implementations for these application scenarios to facilitate the pattern application to retrieve an executable application.

Reinfurt~et~al.~\cite{Reinfurt2019_EntryPoints} present an approach for identifying entry points in pattern languages.
To support practitioners in finding suitable entry points for their problems, they first assess the situation, i.e., the problem and requirements.
Then an entry point and the shortest path through the pattern language solving the described situation is searched, and finally, the patterns along this path are applied.
In contrast to their approach, our method utilizes the identified entry point patterns to build a graph consisting of concrete solutions enabling an automated aggregation of these concrete solutions to build a hybrid quantum application.

Various papers focus on the usage of machine learning and text matching to represent textual input as graphs comprising all important information.
Osman~et~al.~\cite{osman2020graph} summarize different techniques to represent text as graphs and how the graphs can be generated and analyzed.
Castillo~et~al.~\cite{castillo2017text} present concepts to analyze text documents from the web and extract the relevant features, which are then represented as graphs. 
Our approach includes these concepts to generate pattern graphs based on given textual input.\looseness=-1

Misra~et~al.~\cite{misra2021build} introduce an interactive advisor to support users in deciding if quantum computing is suitable for a given problem and the following design decisions when realizing a quantum application.
For this, the user provides a textual description of the problem, which is then automatically analyzed to identify possible quantum algorithms to use utilizing natural language similarity analysis.
However, in contrast to our approach, it is not based on patterns providing the user additional information by highlighting the problems to solve with proven solutions, and does not support automatically building a corresponding hybrid quantum application.\looseness=-1

Different frameworks enable the generation of quantum circuits or quantum applications based on \textit{Model-Driven Engineering (MDE)} techniques.
A research roadmap for model-driven quantum software engineering is presented by Gemeinhardt~et~al.~\cite{gemeinhardt2021towards}.
They present different potential research directions and open research questions to support the generation of hybrid quantum applications from models.
QPath~\cite{qpath2024} presents a toolchain aiming at supporting all phases of the quantum software lifecycle from requirement analysis, through architecture and implementation, to the operation of the application.
Classiq~\cite{minerbi2022quantum} introduces a platform to automatically generate large-scale quantum circuits based on given requirements and constraints.
They specifically focus on larger quantum circuits that are very complex to realize manually.
P{\'e}rez-Castillo~et~al.~\cite{perez2023generation} present an approach to generate quantum circuits using model transformations.
For this, they use UML to define their input models.
A framework to abstractly define quantum circuits is developed by Gemeinhardt~et~al.~\cite{gemeinhardt2024model}.
They utilize a novel modeling language that is based on so-called composite operators.
These operators abstract technical and quantum-specific details and ease the design process.
Finally, the framework enables the generation of executable quantum circuits.
However, all these approaches only support the generation of quantum circuits and not whole quantum applications.\looseness=-1

\section{Conclusion \& Outlook}
\label{sec:fw}
\noindent
To reduce the complexity when developing hybrid quantum applications, we presented an approach to automatically identify suitable quantum computing patterns for building a hybrid quantum application based on textual user input describing the problem at hand as well as non-functional requirements.
Moreover, the approach enables selecting concrete solutions corresponding to the identified patterns that satisfy the non-functional requirements, e.g., privacy or runtime constraints.
If suitable concrete solutions as well as the required aggregation operators, describing how the solutions can be combined, are available, the concrete solutions can be automatically aggregated to a deployable and executable hybrid quantum application.
To demonstrate the practical feasibility of our approach, we introduced a system architecture, a prototypical implementation, and a case study generating an application using Grover`s algorithm to solve the 3-SAT problem.\looseness=-1

In future work, we plan to evaluate how effectively our approach can support partners from industry and academia in solving their quantum computing use cases.
Based on these evaluation results, we plan to iteratively adapt our prototypical implementation increasing its practical applicability for real-world use cases.
Furthermore, we will identify additional patterns to extend the quantum computing pattern language and incorporate them into our approach.

\section*{Acknowledgements}

This work was partially funded by the BMWK projects \textit{EniQmA}~(01MQ22007B) and \textit{SeQuenC}~(01MQ22009B).

\footnotesize
\bibliographystyle{IEEEtranN}
\bibliography{main}

All links were last followed on \today.

\end{document}